# Towards a novel laser-driven method of exotic nuclei extraction-acceleration for fundamental physics and technology

M. Nishiuchi[1], H. Sakaki[1], K. Nishio[2], R. Orlandi[2], H. Sako[2,3], T. A. Pikuz[1,6], A. Ya. Faenov[1,6], T. Zh. Esirkepov[1], A. S. Pirozhkov[1], K. Matsukawa[1,4], A. Sagisaka[1], K. Ogura[1], M. Kanasaki[1,4], H. Kiriyama[1], Y. Fukuda[1], H. Koura[2], M. Kando[1], T. Yamauchi[4], Y. Watanabe[5], S. V. Bulanov[1], K. Kondo[1], K. Imai[2] and S. Nagamiya[7]

[1]*Kansai Photon Science Institute, Japan Atomic Energy Agency, Kizugawa, Kyoto, Japan*

[2]*Advanced Science Research Center, Japan Atomic Energy Agency, Tokai, Ibaraki, Japan*

[3]*J-PARC Center, Tokai, Ibaraki, Japan*

[4]*Graduate School of Maritime Sciences, Kobe University, Japan*

[5]*Interdisciplinary Graduate School of Engineering Sciences, Kyushu University, Japan*

[6] *Joint Institute for High Temperature of RAS, Moscow, Russia*

[7]*RIKEN, 2-1 Hirosawa, Wako, Saitama, Japan*

**The measurement of properties of exotic nuclei, essential for fundamental nuclear physics, now confronts a formidable challenge for contemporary radiofrequency accelerator technology. A promising option can be found in the combination of state-of-the-art high-intensity short pulse laser system and nuclear measurement techniques. We propose a novel *Laser-driven Exotic Nuclei extraction-acceleration method* (LENex): a femtosecond petawatt laser, irradiating a target bombarded by an external ion beam, extracts from the target and accelerates to few GeV highly-charged nuclear reaction products. Here a proof-of-principle experiment of LENex is presented: a few hundred-terawatt laser focused onto an aluminum foil, with a small amount of iron simulating nuclear reaction products, extracts almost fully stripped iron nuclei and accelerate them up to 0.9 GeV. Our experiments and numerical simulations show that short-lived, heavy exotic nuclei, with a much larger charge-to-mass ratio than in conventional technology, can be obtained in the form of an energetic, low-emittance, high-current beam.**

**[151 words]**

The measurement of the properties of exotic nuclei (e.g., decay and capture rates), especially for the short-lived ones, is necessary for understanding nucleosynthesis governing the dynamics of stellar objects e.g. novae, supernovae, and x-ray bursters[1]. In such celestial bodies, consecutive rapid neutron captures create almost all chemical elements heavier than iron indispensable for our life, such as copper, zinc or iodine. It is difficult to determine how the nucleosynthesis is influenced by short-lived neutron-rich actinides and super-heavy nuclei (transactinides) because their properties are largely unknown[2].

State-of-the-art *radio-isotope sources*[3] enable the exploration of exotic nuclei beyond the so-called *Heisenberg valley*[4]. This matured technology, however, is reaching its limits on the production of heavy, neutron-rich nuclei, due to their extremely low production cross sections and very short half-lives. The simultaneous creation of a large amount of unwanted isotopes aggravates the problem. A novel type *radio-isotope source* is needed to explore the frontiers of the nuclear chart. The new source must have a high production rate, fast extraction time, sufficient selectivity, and adaptability to the contemporary measurement techniques.

At present, *radio isotope sources* are mainly based on two methods: *isotope separation on-line* (ISOL)[5] and *projectile fragmentation* (PF)[6], exemplified by CERN's ISOLDE[7] and RIKEN's RIBF[8], respectively. In the ISOL method, unsuitable for measurements of nuclei with life-times less than few hundred milliseconds, the isotope extraction strongly depends on the target chemical properties of the isotope source, and in some cases the extraction efficiency is less than a percent[9]. For the measurement of short-lived isotope properties, often ISOL requires facilities for ionization and acceleration. The PF method, suitable for nuclei with lifetimes down to 100 ns, produces an isotope beam with an undesirably large emittance[4] of few tens of $\pi$·mm·mrad. With PF, isotopes heavier than the projectile cannot easily be studied, and only via secondary reactions[10].

Laser-driven ion acceleration can extract from the target exotic nuclei with short lifetime, including transactinides, for which both above-mentioned methods face difficulties. Present-day petawatt-class laser[11,12] produces femtosecond pulses focusable to a few-microns spot reaching the intensity of $10^{22}$ W/cm$^2$. Interacting with various targets these laser pulses generate quasi-static electric fields with the strength of ~100TV/m which can efficiently strip atoms and accelerate the resulting ions within sub-micron meter distance at energies up to hundreds of MeV per nucleon (MeV/u). The obtained ion beams show a low transverse emittances of $10^{-4}$ $\pi$·mm·mrad[13]. Several mechanisms of ion acceleration were identified[12]. Heavy ion acceleration up to few

MeV/u energies has been observed experimentally with a 400 joule sub-picosecond laser pulses[14]. Numerical simulations show that a petawatt laser irradiating a compound target with additional charge stripper can provide a heavy ion injector for conventional accelerators[15].

In this paper we propose a novel *Laser-driven Exotic Nuclei extraction-acceleration method* (LENex), summarized schematically in Fig.1a (see also Supplementary Information SI_1). An external ion beam supplied by a conventional accelerator is used to produce radioactive isotopes which are stopped in a micron-thick solid target. A petawatt laser pulse extracts and accelerates the reaction products irrespective of the target chemical properties. The resulting beam of almost fully stripped ions, with a much larger charge-to-mass (Q/M) ratio than in conventional technology[16], is separated by magnets into beamlets with a different Q/M ratio for measurements in different chambers without any need for additional post-acceleration. This method would permit to access a large area of yet-unknown nuclei in the nuclear chart, Fig. 1b.

Here we show the first proof-of-principle experiment of LENex: we demonstrate that a 200TW-class laser pulse with the energy less than 10J, focused to a thin aluminum foil with a small amount of iron simulating the exotic nuclei, generates a GeV iron beam with large Q/M ratio. A feasibility study of the specific example of LENex using numerical simulations is also presented.

## Results

### Proof of principle experiment of LENex

The J-KAREN laser[17], with 0.8 μm wavelength and repetitive operation capability produces 35 fs, 8 J laser pulses which are focused onto a target at a 45° incidence angle, as is shown in Fig. 2a, forming a focal spot of 3 μm. The resulting peak intensity is $10^{21}$ W/cm$^2$. The target is a 0.8-μm-thick aluminium (Al) foil with a few nanometer thick patches of iron (Fe) on the surfaces. The Fe ions, which have larger atomic number than Al, simulate the presence of nuclear reaction products (produced, for example, via fusion or spallation reactions), which cannot be studied using the aforementioned *radio-isotope* source. The Al foil thickness is chosen with respect to the J-KAREN laser parameters in order to maximize the energies of the simultaneously accelerated protons, which originate from water contamination on the surface of the Al foil, by using the results of previous experiments[18]. Those protons can be removed beforehand, if necessary, by already know methods[14,19]. The total amount of iron contained in the target is about 0.5% in terms of number density of atoms (for details see Supplementary Information, Fig.SI_2). The films in the stack detector are used for detection of ions which are accelerated during the laser-foil interaction. These ions leave traces in the films which, via the etching process, are revealed after the experiment as pits whose size differ depending on ion kind and energy. Films with different detection threshold are employed, and Kapton films record the Fe ion tracks. By controlling the etching condition, Fe pits are discriminated from lighter ion pits (see Supplementary Fig.SI_3). From the total number of pits in each Kapton film, it appears that $\gtrsim 10^6$ Fe nuclei per shot are produced, with an energy from 0.56 to 0.89 GeV (corresponding to the interval from 10 to 16 MeV/u), Fig. 2c. Such the rate is already sufficient for nuclear measurement[20], while the energy, surpassing previous experimental results, is suitable for measurements without the necessity of post-acceleration.

The charge state of Fe ions can be inferred from X-ray spectra[21], which exhibit the presence of $Fe^{+25}$ carrying only one electron and $Fe^{+24}$ with two electrons, as shown in Fig. 2b. The ionization occurs in several ways. The optical field ionization mechanism[22] can create $Fe^{+23}$ whose ionization potential is 2.1 keV. Further charge stripping occurs due to electron impact ionization[23] with tens-of-keV electrons which appear in the laser pulse interaction with the target[24] and due to single photon ionization mechanism[23] by laser-plasma produced x-rays[25]. A synergy between these processes creates $Fe^{+25}$ (ionization potential: 8.8 keV) and, most probably, fully stripped iron, $Fe^{+26}$ (ionization potential: 9.3 keV). We expect that under similar conditions uranium atoms (U) can be ionized to $U^{+72}$ since the ionization potential (8.8 keV) is close to that of $Fe^{+25}$.

Using particle-in-cell (PIC) simulations with the code REMP[26] we reproduce the results of our experiments, as shown in Fig.3. In the vicinity of the laser focus, the target is assumed to be fully ionized due to the above-mentioned effects[22,23]. Attached to the back side of the target, a 5 nm thick $Fe^{+26}$ and a 10 nm thick oxygen ($O^{+8}$) and hydrogen ($p^{+}$) layers simulate, respectively, iron patches and water ($H_2O$) contamination present in the experiment. On the laser irradiated side, the target has preplasma created by the prepulse existing due to a laser finite contrast, as predicted by hydrodynamic simulations[27]. Most of the laser pulse energy is absorbed during the interaction with the target, while less than 5% is reflected into the acceptance angle of the reflectivity monitor at the specular direction, as it can be seen in Fig. 3a. On the back side of the target ion species (dominant species, e.g., $p^{+}$, $O^{+8}$, $Al^{+13}$, $Fe^{+26}$) are accelerated by the electrostatic potential induced in the laser-target interaction, as shown in Fig. 3 b, c, d. The amount of laser reflection at the specular direction and the ion spectra are in agreement with the experimental findings.

## Discussion

We demonstrate the feasibility of LENex considering the example of not-yet-discovered neptunium isotope which is at present difficult to extracts with other techniques. Neptunium isotopes can be created in a curium foil by a bombardment of high-energy protons, and later extracted and accelerated by a laser.

A 5 mm diameter beam of 400 MeV protons with the average current of 50 μA bombards a 0.8 μm thick curium ($^{248}$Cm) foil with the areal density of ~1000 μg/cm$^2$. Among different isotopes, neptunium-245 ($^{245}$Np) with the predicted half-life of 2.29 minutes[28], shown in the inset of Fig. 1b, is produced with the cross-sections of $1.6\times10^{-31}$ cm$^{-2}$, as calculated using the EPAX3 model[29] with the code LISE++[30]. A one-second bombardment creates ~$10^2$ of $^{245}$Np. The foil is then irradiated by a 1 Hz, 1 PW laser for extraction. Assuming that 1% of ions are extracted from a 30 μm wide acceleration region, we obtain $4\times10^{-5}$ neptunium-245 nuclei per second. The corresponding event rates are 3.3 of $^{245}$Np per day. Thus $^{245}$Np, inaccessible with conventional techniques, can be obtained with the proposed scheme. Accelerated to 2 MeV/u, the nuclei travel 10 m in 1 μs which is negligible compared to their half-lives. Varying the target composition the laser can extract different exotic nuclei irrespective of their chemical properties.

Aiming at future laser parameters of interest for LENex, the results of multi-parametric PIC simulations[26] of an idealized laser pulse interaction with a pure curium ($^{248}$Cm) foil can be examined. Varying the laser pulse energy and the foil thickness we find the maximum energy and number of accelerated $^{248}$Cm ions with different Q/M ratios (Fig. 4, Supplementary Information Fig.SI_4-6). For example, even with the currently achievable parameters by the sub-petawatt-class laser of, 35 J of energy, 35 fs of the duration, and a 3 μm wide focal spot, can extract more than $5\times10^{10}$ ions with an energy above 1 MeV/u from a 32 μm wide area of a 0.8 μm thick curium foil. This gives about 1% ion extraction efficiency used in the above estimation. The maximum ion energy (117 MeV/u) is reached for the optimal target thickness, (see the Supplementary Information Fig.SI_7). We note that the conclusions are almost the same for isotopes with a close to $^{248}$Cm atomic weight, e.g., the above-considered $^{245}$Np, provided that the Q/M ratio is the same.

A more compact facility can emerge if the external ion beam from a conventional driver is replaced by a laser-driven source with a future multi-petawatt laser. This alternative setup would exploit one of the most efficient acceleration mechanisms, the *radiation pressure dominant acceleration* (RPDA)[31], which can extract and accelerate nearly 100% of ions located in the laser focal spot. Among those ions, exotic short-lived nuclei acquiring relativistic velocity are transported to large distances without significant losses due to relativistic time dilation, facilitating the measurement of their properties.

In conclusion, we have experimentally demonstrated that the GeV energy beam of almost fully-stripped charge state heavy ions, simulating nuclear reaction products, are extracted and accelerated from the micron-thick foil target by the 200-terawatt laser with an energy less than 10 J. Using numerical simulations of a 1Hz petawatt laser irradiation with a micron-thick curium foil, and of a 400-MeV proton beam collision with curium, we have shown that the laser can extract and accelerate to more than 1MeV/u about $5\times10^{10}$ ions, which contain 3.3 nuclei of $^{245}$Np isotope per day. Thus, LENex can enable the measurements of this not-yet-discovered element properties.

A combination of state-of-the-art techniques of nuclear measurements and the petawatt laser-driven ion acceleration opens up new opportunities in the experimental nuclear physics. The proposed LENex scheme, which is conceptually new, can be widely adopted in the existing radio-isotope facilities, since it gives access to a large area of nuclei in the nuclear chart. It will enable measurements of the properties of exotic nuclei at present inaccessible with conventional radio isotope facilities and heavy ion accelerators. The proven laser-driven acceleration of a GeV beam of almost fully-stripped heavy Fe ions also expedites the realization of a compact laser-driven heavy ion accelerator. These consequences will have far-reaching implications in different fields of science and technology.

**[1980words]**

**Methods**

**Laser.** The experiments are performed using the J-KAREN[17], an optical parametric chirped-pulse amplification (OPCPA) Ti:sapphire hybrid laser system at Kansai Photon Science Institute of Japan Atomic Energy Agency. Laser pulses have the energy of 8 J, duration of 35 fs (full-width-at-half-maximum, FWHM, with respect to intensity), wavelength of 0.8 μm and a typical contrast of $10^{10}$ (with respect to power; at several picoseconds before the pulse peak[17]). The contrast is achieved with a saturable absorber inserted between the high-energy CPA oscillator and stretcher. Using F/2.1 off-axis parabolic mirror, p-polarized laser pulses are focused onto a target at an incidence angle of 45°, producing a spot with the FWHM diameter of 3μm (FWHM) and the peak laser intensity of $10^{21}$ W/cm$^2$ [18]. The peak intensity estimation is consistent with a typical 10 MeV slope temperature of electron energy spectra measured by the magnetic electron spectrometer installed along the laser propagation axis. Preplasma due to finite contrast effects is monitored by observing the laser reflectivity from the target within ~1 radian at specular direction.

**Target.**

A 0.8 μm thick Al foil is placed at the laser focus position with a 5 μm accuracy along the laser propagation axis, well within the Rayleigh length of 20 μm. Al target contains small amount of Fe which simulates the exotic nuclei produced by nuclear reactions. See the Supplementary Information for further details.

**Diagnostics for Ion and X-ray spectra**

The stack detector includes Kapton films used for the detection of iron ions and calibrated with the ions from conventional accelerator, the HIMAC at NIRS in Chiba, Japan[32]. Supplementary information contains details on the procedure of obtaining ion spectra.

The x-ray spectra are obtained using the focusing spectrometer with spatial resolution (FSSR) equipped with a spherically bent mica crystal and a back illuminated CCD camera[21]. In the aluminum target with 0.5% of iron impurity, the x-ray signal peculiar to Fe ions is below the detection threshold. Therefore we use a 2-μm-thick iron (stainless steel) foil irradiated by the same laser. The resulting x-ray spectrum indicates the presence of $Fe^{+25}$ ions Fig. 1b. Since the difference in ionization potentials of $Fe^{+25}$ and $Fe^{+26}$ is small, the production of fully stripped iron ions is highly probable. We note that $Fe^{+24}$ lines accompany many well-known satellite lines[33].

**Particle in cell (PIC) simulation code**

The REMP is relativistic electromagnetic multi-dimensional and multi-parametric[26] particle-in-cell code suitable for simulations of intense laser-matter interactions. Supplementary information contains details on the simulation setup and results.

**LISE++.**

The LISE++ code predicts the intensity and purity of radioactive ion beams produced by in-flight separator. The cross-sections of different reactions resulted from the interaction of 400-MeV proton with $^{248}Cm$ nucleus are calculated using EPAX3 model[28] implemented in the program LISE++[29].

**[452 words]**

**Acknowledgements**
The authors would like to acknowledge and thank the J-KAREN laser operation team for continuous support throughout the experimental campaigns. We would like to thank Keitetsu Kondo and Eiichi Wakai for carrying out the SEM analysis of the target material and for the useful discussions. We would like to thank Noboru Hasegawa and Tetsuya Kawachi for supporting the measurement of X-ray spectra. This work was supported by JSPS KAKENHI (Grant No. 24740280 and Grant No. 25390135) .

**Author contributions**
M. Nishiuchi, H. Sakaki, T. A. Pikuz, A. Ya. Faenov, A. S. Pirozhkov, K. Matsukawa, A. Sagisaka, K. Ogura, and M. Kanasaki performed experiments using J-KAREN laser; T. Zh. Eirkepov performed simulations with code REMP; K. Nishio, H. Sako, R. Orlandi, H. Koura, and K. Imai performed and interpreted simulations with code LISE++; in other aspects all authors contributed equally to this work.

**Additional information**
Supplementary information is available in the online version of the paper. Reprints and permissions information is available at www.nature.com/reprints. Correspondence and requests for materials should be addressed to M.N.

**Competing financial interests**
The authors declare no competing financial interests.

**Figure legend**

**Fig.1. The novel laser-driven exotic nuclei extraction-acceleration method (LENex), and its possible accessibility region in the nuclear chart.**

**a** The proposed setup scheme. A target containing material A is bombarded by ions (e.g. protons) with sufficient energy to produce the isotopes of interest B, C, etc. A petawatt-class laser focused onto the target accelerates the new isotopes to an energy of few GeV. After theseparation, isotopes are measured in different experimental chambers.

**b** Nuclear table colored according to their half-life [NIST Atomic Spectra Database, http://www.nist.gov/pml/data/asd.cfm]. The horizontal and vertical axes, respectively, represent the neutron and proton number in a nucleus. The proposed laser-driven source provides opportunities to measure exotic nuclei in the orange zone. As one of the examples of feasibility of obtaining an exotic nuclear beam, the case of $^{245}Np$ is presented in the text.
**[117 words]**

**Fig. 2. Experimental set-up with the J-KAREN laser, target, diagnostics and obtained data.**

**a** The experiment scheme. A 8 J, 35 fs generated by the J-KAREN laser is focused onto a thin foil target with the peak intensity of $10^{21}$ W/cm$^2$. X-rays from the laser-foil interaction region are measured by a focusing spectrometer with spatial resolution (FSSR)[20] equipped with a spherically bent mica crystal and a back illuminated CCD camera. Ions accelerated in the laser-foil interaction are measured by a stack detector.

**b** Typical X-ray spectrum in the case of a 2 μm thick iron (Fe) foil target in the spectral range of 1.70 – 1.97 [Å]. Distinct peaks (called spectral lines) in red correspond to emission from almost fully stripped iron ions in the charge state of $Fe^{+25}$ and $Fe^{+24}$.

**c** The energy spectrum of iron ions accelerated from a 0.8 μm thick aluminium (Al) foil with a few nanometer thick iron patches, as measured with the Kapton layers in the stack detector. Supplementary Information contains details on the target and the procedure of Fe ion detection.
**[164 words]**

**Fig. 3. PIC simulation results supporting experimental data.**

A 8 J, 35 fs laser pulse interacts with a target comprising a 0.8 μm thick Al foil, a 5 nm thick Fe and 10 nm thick $H_2O$ layers at the rear and preplasma at the front. Laser pulse reflection in terms of dimensionless electric field strength (**a**), ion species acceleration revealed by the ion density (**b,c**) and the ion energy spectra (**d**). Supplementary Information contains details on the simulation setup.

**[71 words]**

**Fig. 4. PIC simulation results showing the feasibility of LENex.**

The maximum energy of Curium ($^{248}Cm$) ions, $\mathcal{E}_{\mathrm{Cm}}$ [MeV/u], vs the laser pulse energy, $\mathcal{E}_{las}$ [J], target thickness, $\ell$ [μm], and ion Q/M ratio (normalized to that of proton). Surfaces corresponds to $\mathcal{E}_{\mathrm{Cm}}$ =2, 6.5, 30, 70, 90, and 100 [MeV/u]. Color encodes the number of accelerated ions in all charge states having energy not less than $\mathcal{E}_{\mathrm{Cm}}$. Supplementary Information contains details on the simulation setup.
**[66 words]**

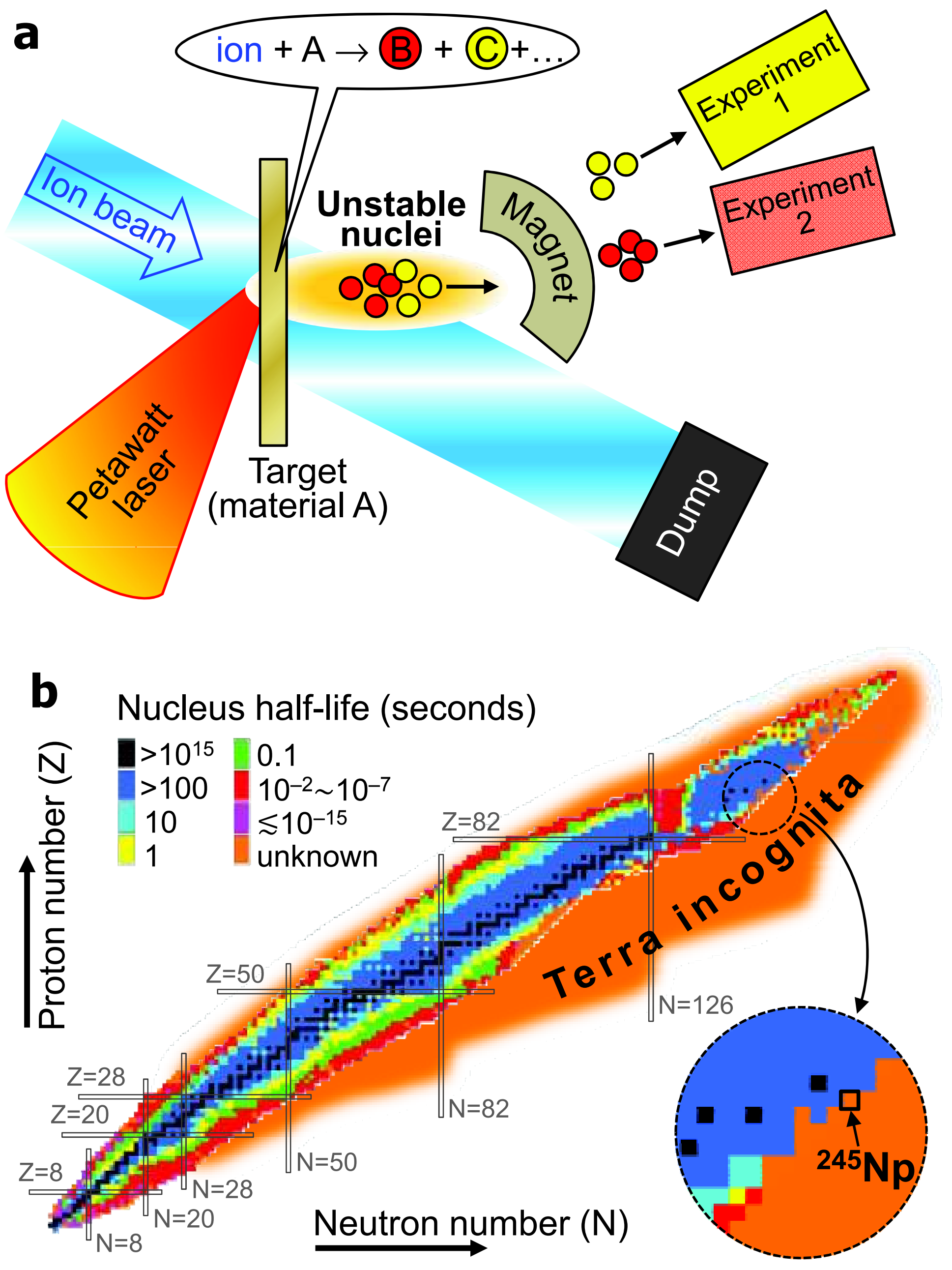
a
ion + A → B + C +…
Ion beam
Petawatt laser
Target (material A)
Unstable nuclei
Magnet
Experiment 1
Experiment 2
Dump
b
Nucleus half-life (seconds)
>10^15
>100
10
1
0.1
10^−2~10^−7
≲10^−15
unknown
Proton number (Z)
Neutron number (N)
Z=8
Z=20
Z=28
Z=50
Z=82
N=8
N=20
N=28
N=50
N=82
N=126
Terra incognita
245Np

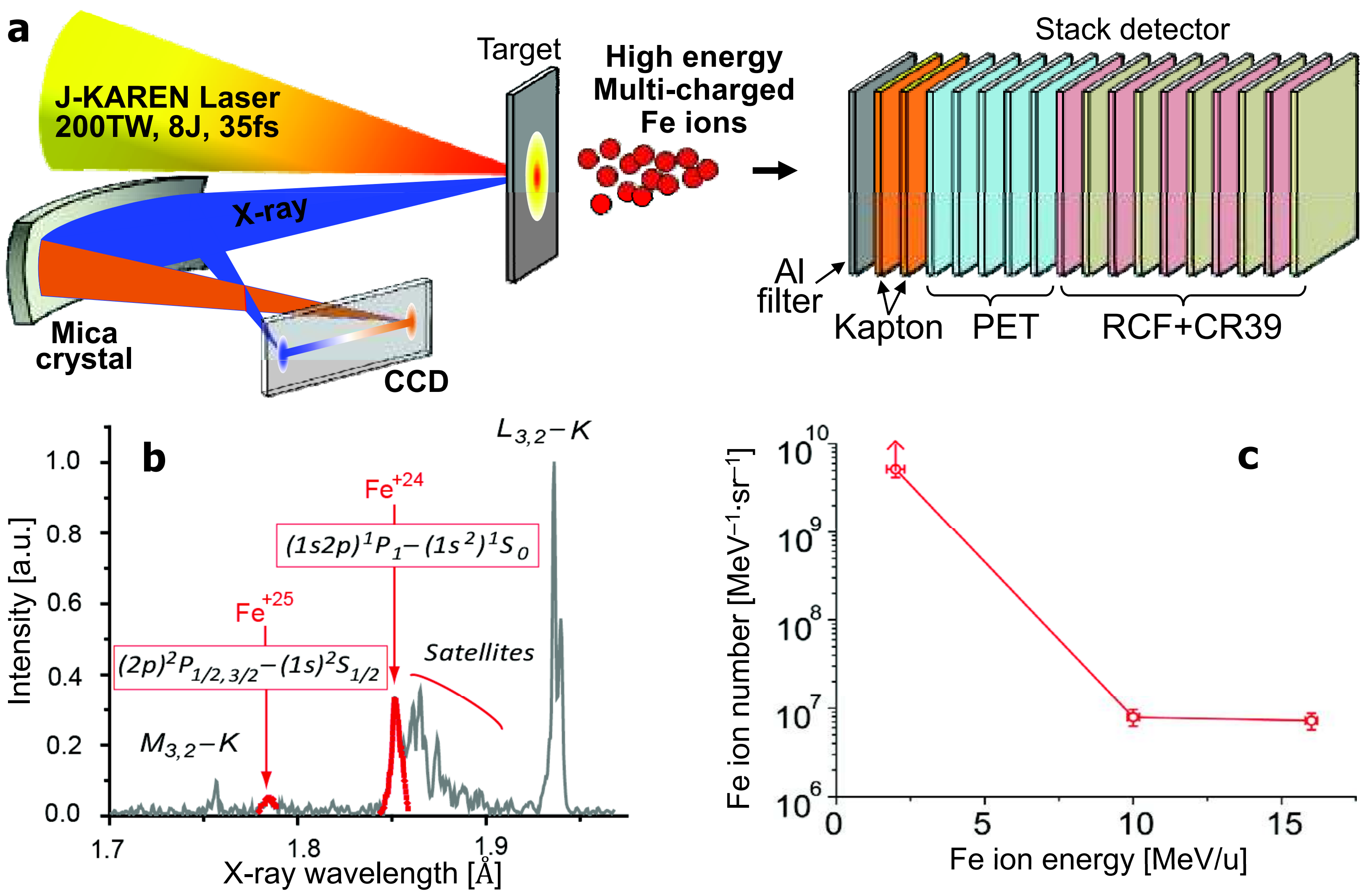

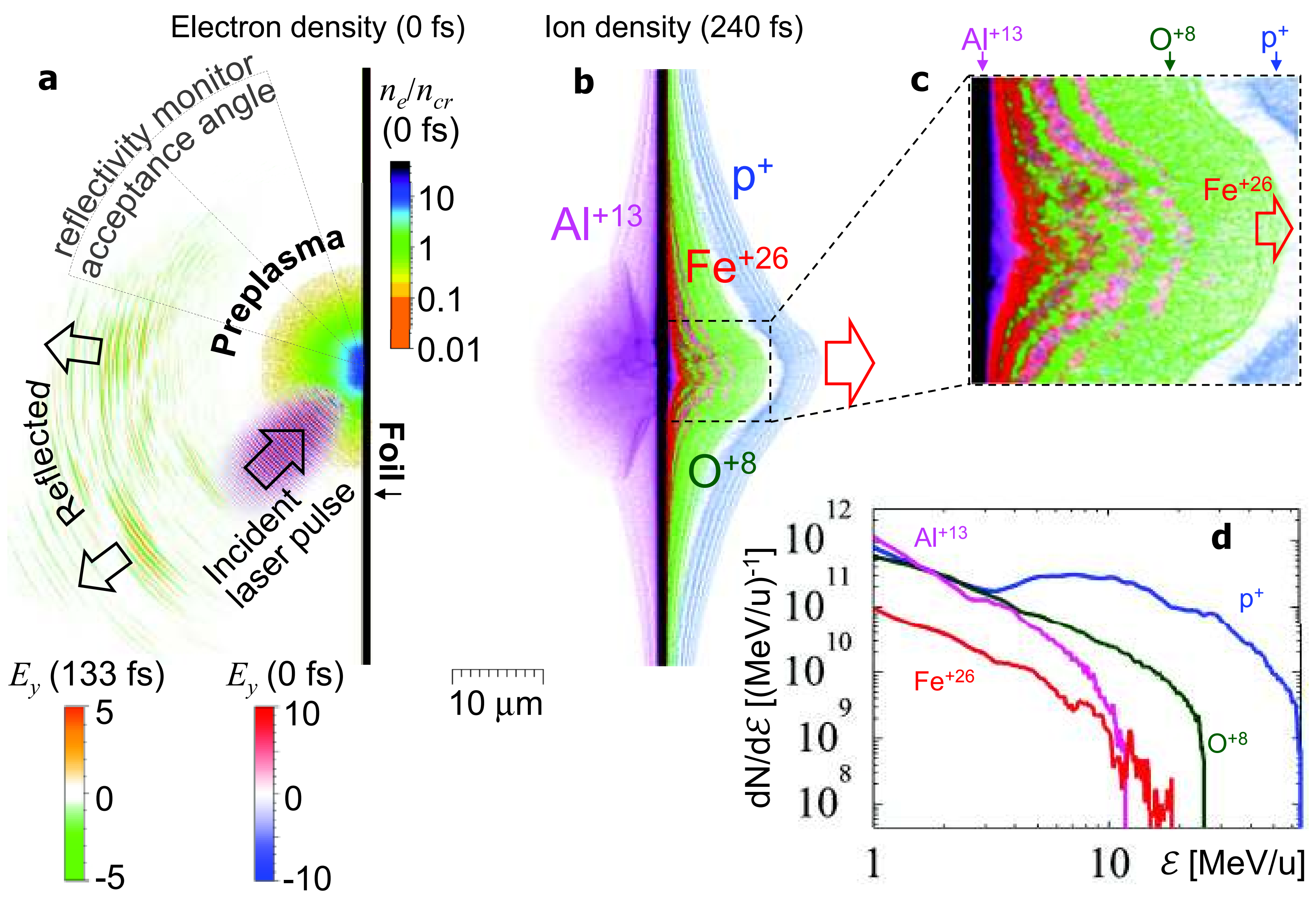

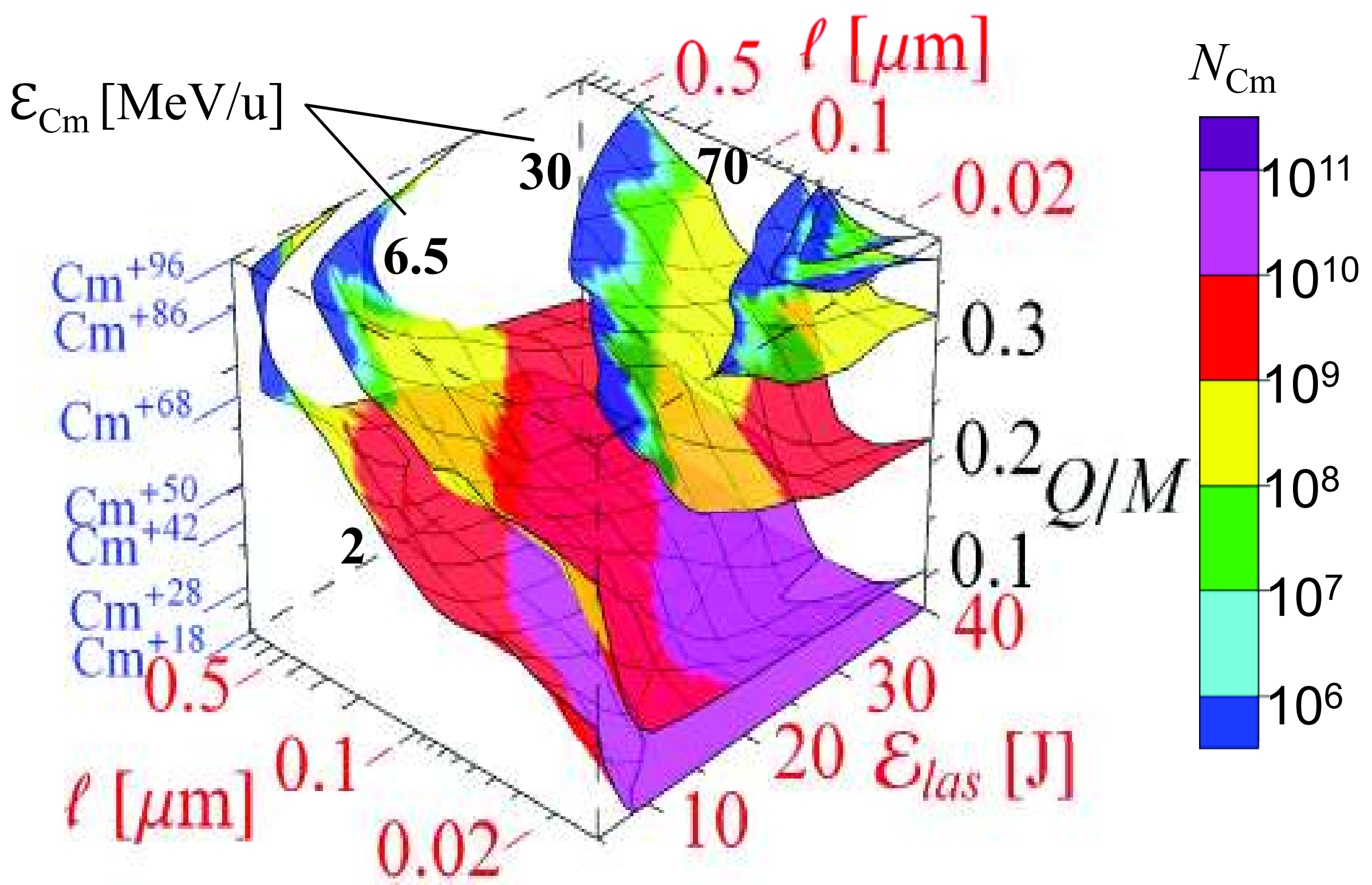

# Towards a novel laser-driven method of exotic nuclei extraction-acceleration for fundamental physics and technology

M. Nishiuchi[1]*, H. Sakaki[1], K. Nishio[2], R. Orlandi[2], H. Sako[2,3], T. A. Pikuz[1,6], A. Ya. Faenov[1,6], T. Zh. Esirkepov[1], A. S. Pirozhkov[1], K. Matsukawa[1,4], A. Sagisaka[1], K. Ogura[1], M. Kanasaki[1,4], H. Kiriyama[1], Y. Fukuda[1], H. Koura[2], M. Kando[1], T. Yamauchi[4], Y. Watanabe[5], S. V. Bulanov[1], K. Kondo[1], K. Imai[2], and S. Nagamiya[7]

*[1]Kansai Photon Science Institute, Japan Atomic Energy Agency, Kizugawa, Kyoto, Japan*

*[2]Advanced Science Research Center, Japan Atomic Energy Agency, Tokai, Ibaraki, Japan*

*[3]J-PARC Center, Tokai, Ibaraki, Japan*

*[4]Graduate School of Maritime Sciences, Kobe University, Japan*

*[5]Interdisciplinary Graduate School of Engineering Sciences, Kyushu University, Japan*

*[6]Joint Institute for High Temperature of RAS, Moscow, Russia*

*[7]RIKEN, 2-1 Hirosawa, Wako, Saitama 351-0198, Japan*

*e-mail: nishiuchi.mamiko@jaea.go.jp

Here we present Supplementary Information. All acronyms, annotations, etc. and references are the same as defined in the main manuscript, if not stated otherwise. The material is organized in few sections as follows;

1. **Target used in the experiment;**
2. **Procedure of obtaining Fe ion spectrum from stack detector;**
3. **Explanation of PIC simulation presented in Fig.3;**
4. **Explanation of PIC simulation presented in Fig.4;**
5. **Synopsis of the animation SI_1 (Novel laser-driven exotic nuclei extraction-acceleration method);**

**References;**

**Figure legends (SI_2 – SI_7)**

## 1. Target used in the experiment

A 0.8 μm thick aluminum (Al) foil with a small amount of iron (Fe) simulating nuclear reaction products, is used as the target. The scanning electron microscope images of the Al foil target is shown in Fig. SI_2a. The surface quality is revealed by reflected electrons, Fig. SI_1a, and aluminum (Al) Kα spectral line emission, Fig. SI_2b. Iron (Fe) Kα spectral line emission, Fig. SI_2c, show how iron is distributed. The content of different species in the 0.8 μm thick aluminum foil is seen in Fig. SI_2d. These species present on the foil as a few nanometer thick patches. The iron content is 0.5%, orders of magnitude higher than other species which are detectable by Kapton film in our condition of measurement. In addition, 10 nm thick water ($H_2O$) contamination layers usually present on both sides of the foil.

**2. Procedure in obtaining Fe ion spectrum from stack detector.**

For the detection of the ion spectra stack detector is placed behind the target at 55 mm along the target normal. It consists of solid state nuclear track detectors with different detection thresholds: 2 pieces of 125 μm thick Kapton film (made by Nilaco Corporation) and 5 pieces of 100 μm thick Polyethylene terephthalate (PET) films (made by TOYOBO Co. Ltd.) and 15 pairs of 231 μm thick Gafchromic films (RCF: XR-RV3, made by ISP Ltd.) and 900 μm-thick CR-39 (TD-1, made by Nagase Landauer Ltd.) films. The CR-39 can record protons as etchable tracks with energy less than 20 MeV[34]. The PET is insensitive to protons and He ions[35]. The Kapton is insensitive to Ne and lighter ions[36]. The stack detector is shielded by a 12 μm thick Al foil from the direct laser exposure. Protons, dominating ion species due to water contamination, do not affect the heavy ions etch-pits recorded by Kapton and PET films. Hereafter we concentrate on the data recorded only on Kapton films and those recorded on PET and CR-39 films are reported elsewhere. The ion energy is calculated basing on the depth of the detected ion inside the stack detector using PHITS code[37]

Iron ions are detected by Kapton films which are etched with the sodium hypochlorite solution kept at 55°C. The initial content of active chlorine is between 8.5 and 13.5%. The etching condition is optimized for the detection of ions heavier than argon (Ar), thus even the Al ions from the substrate and Si impurity, which is the second after Fe in terms of content (Fig. SI_2), are unable to make pits with sizes measureable with the optical microscopy. The Kapton film response is calibrated using a 40 MeV Ar ion beam and 140 MeV and 80 MeV Fe ion beams from the conventional accelerator HIMAC in Chiba, Japan. The growth rate of the pit size in the etching process linearly depends on the bulk etch rate, which is defined as the bulk thickness removed in the etching process (Fig. SI_3). The ions with energy corresponding to the Bragg peak reached at the film make pits with larger sizes than ions with greater energies. At the bulk etch of $(4.4 \pm 0.5)$ μm, the pit number is counted and each pit diameter is measured using microscope-based apparatus. Pits with sizes between 1.5 μm (corresponding to 40 MeV Ar) and 2.4 μm (80 MeV Fe) are attributed to Fe ion bombardments. Counting the etch pits recorded on the successive Kapton films (front of 1$^{st}$ and 2$^{nd}$ , and rear side of 2nd films) with the accuracy of ~1%, we reconstruct the energy spectrum of Fe ions, shown in Fig. 2c. The number of etch pits recorded at the front side of the first Kapton film exceeds the saturation level.

**3. Explanation of PIC simulation presented in Fig.3.**

Using REMP code we were able to reproduce the results of our experiments. The laser prepulse existing due to a finite contrast creates preplasma, whose properties were predicted by hydrodynamic simulations[27]. Preplasma is inserted as an initial condition into PIC simulation, where a p-polarized Gaussian laser pulse with the energy of 8 J and the FWHM duration of 35 fs is incident onto a 0.8 μm thick aluminum (Al) target at 45°. The laser focal spot is 3 μm (FWHM with respect to intensity) in vacuum. The plasma is assumed to be fully ionized due to the optical field ionization[22], electron impact ionization and x-ray photoionization[23]; the scalelength of preplasma is 4.8 μm (electron density drops 10 times on that distance), the critical surface is located ~6 μm from the solid density level. The target also contains a 5 nm thick iron (Fe) layer and 10 nm water ($H_2O$) contamination layer on the back side. The simulation box is 104 μm × 80 μm, mesh size is 25 nm, number of quasiparticles is $1.5\times10^9$.

About 75% of the laser pulse energy is absorbed during the interaction with the target, while less than 5% is reflected into the acceptance angle of the reflectivity monitor[38] at the specular direction, Fig. 2 a. On the back side of the target ion species ($p^+$, $O^{+8}$, $Al^{+13}$, $Fe^{+26}$) are accelerated by the electrostatic potential induced in the laser-target interaction Fig. 2b, c, d. The amount of laser reflection at the specular direction and the ion spectra are in agreement with experimental findings.

In the simulation, we observe that Fe ions behave as probe particles not affecting the accelerating potential, owing to the fact that the number of Fe ions corresponding to the experimental condition (about a 0.5% content of iron) is small. Being accelerated by an external quasistatic electric field, ions with higher Q/M ratio acquire higher energy. For $Fe^{+24}$, $Fe^{+25}$, and $Fe^{+26}$ the energy difference is 5-8%, insignificant for the interpretation both the experiment and simulation results. Therefore the agreement with the experiment strongly supports that detected Fe ions are almost fully stripped.

**4. Explanation of PIC simulation presented in Fig.4.**

Aiming at future laser parameters of interest for the proposed scheme, we performed multi-parametric PIC simulations[26] with an idealized laser pulse and a foil target. A finite laser contrast can modify the interaction, e. g., by causing preplasma formation which can increase the ion energy[27]. However, idealized model is useful for references and gives a prediction for highly controllable laser-plasma interaction which can be accessible with future lasers. The laser pulse is p-polarized and Gaussian, its FWHM duration is 30 fs, its focal spot is 3 μm (the corresponding f-number is 3). The laser pulse energy, $\mathcal{E}_{las}$, varies from 5 to 40 J. The laser pulse is normally incident on the target. The target is Curium ($^{248}$Cm) foil with the density of $3.3\times10^{22}$ atoms per $cm^3$ and the thickness, $\ell$, varying from 12.5 nm to 0.8 μm. We model the result of various mechanisms of ionization[22, 23] by the following distribution of curium charge states: for one part of $Cm^{+18}$ (easily achievable due to OFI[22]) we have 0.18 parts of $Cm^{+28}$, 0.1 of $Cm^{+42}$, 0.056 of $Cm^{+50}$, 0.03 of $Cm^{+68}$, 0.018 of $Cm^{+86}$, and 0.01 of $Cm^{+96}$. The ion charge-to-mass ratio, Q/M, is normalized to the proton charge-to-mass ratio, e. g. for $Cm^{+18}$: Q/M=18/248.

In the course of the laser-target interaction, the ions are accelerated forward, in the direction of the laser pulse propagation. For each particular case of the laser pulse energy and the foil thickness we obtain maximum ion energies, $\mathcal{E}_{\rm Cm}$, corresponding to different Q/M ratios (Fig. SI_4) and the number of ions accelerated to an energy above certain threshold (Fig. SI_5). The resulting function of 3 parameters, $\mathcal{E}_{\rm Cm}(\mathcal{E}_{las}, \ell, {\rm Q/M})$, is presented in Fig. 4, were we show several surfaces of equal energy colored according to the number of ions with the same or greater energy, Fig.SI_4-6. Ions with energy $\gtrsim$1 MeV/u can be relatively easily extracted for subsequent beam manipulation and measurement of nuclei properties. At the threshold of ~6.5 MeV/u $^{248}$Cm nucleus is capable to overcome the Coulomb barrier when colliding with another $^{248}$Cm nucleus. The maximum ion energy (117 MeV/u) is reached for the maximum laser pulse energy and the optimal target thickness, when the foil becomes transparent for the laser pulse at the onset of the RPDA[31], Fig. SI_7.

## 5. Synopsis of animation SI_1 (Novel laser-driven exotic nuclei extraction-acceleration method)

a) A petawatt laser is comparable by size with the state-of-the-art synchrotron accelerators, but the interaction region is just a few cubic centimeter.
b) In the interaction chamber, a thin actinide foil target is bombarded by an external ion beam (e. g. protons or heavy ions) with the energy of a few hundreds MeV/u produced by a conventional accelerator or another laser.
c) The resulting nuclear reactions create new isotopes within the foil target.

d) The petawatt laser pulse is focused onto the foil modified by the bombardment.
e) New isotopes together with other ions from the foil acquire a high charge-to-mass ratio and gain the energy of hundreds of MeV in the course of laser-target interaction.
f) The accelerated highly charged isotopes are separated by beam optics (primarily based on magnets) and used for measurements in different experimental chambers.
g) New data obtained by this method will be crucial for fundamental physics, in particular, for nuclear astrophysics.

---------

**-----**

**Figure legends**

**Fig. SI_2 Target properties.**

Images of the 0.8 μm thick Al foil surface produced with scanning electron microscope by

a. reflected electrons,
b. aluminium (Al) Kα emission, and
c. iron (Fe) Kα emission.
and
d. Different species contamination in the 0.8 μm thick Al foil (parts per million,

in terms of atoms).

**Fig. SI_3 The Kapton film response.**

The Kapton film response is calibrated using a 40 MeV Ar ion beam and 140 MeV and 80 MeV Fe ion beams from the conventional accelerator HIMAC in Chiba, Japan. The growth rate of the pit size in the etching process linearly depends on the bulk etch rate, which is defined as the bulk thickness removed in the etching process.

**Fig. SI_4 Results of multi-parametric simulations(1).**

Maximum ion energies, $\mathcal{E}_{Cm}$, corresponding to different Q/M ratios. Idealized laser pulse and a foil target are used.

**Fig. SI_5 Results of multi-parametric simulations(2).**

The number of ions accelerated to an energy above certain threshold. Idealized laser pulse and a foil target are used.

**Fig. SI_6 Results of multi-parametric simulations(3).**

Laser pulse interaction with a curium-248 foil in the case of laser pulse energy of 35 J and foil thickness of 0.8μm.

**Fig. SI_7 Result of multi-parametric simulations(4).**

Laser pulse interaction with a curium-248 foil in the case of laser pulse energy of 40 J and foil thickness of 25nm.

Fig.SI_2

Images of the 0.8 μm thick Al foil surface produced with scanning electron microscope by

(a) reflected electrons

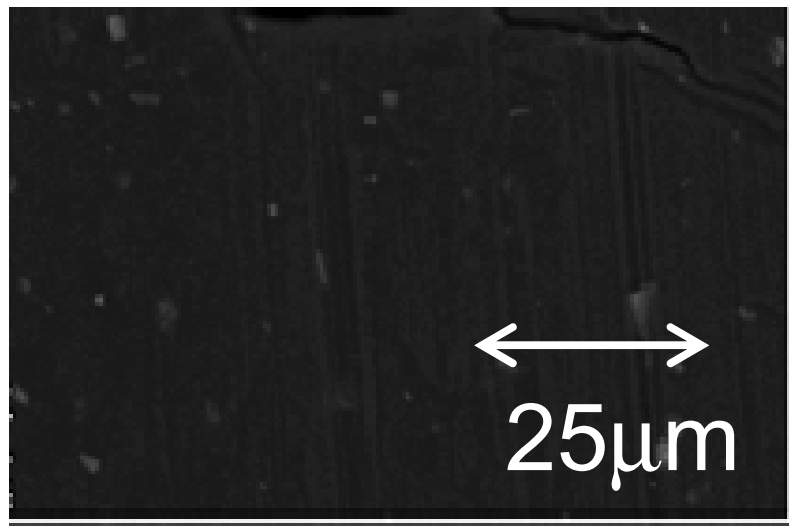

(b) aluminium (Al) Kα emission

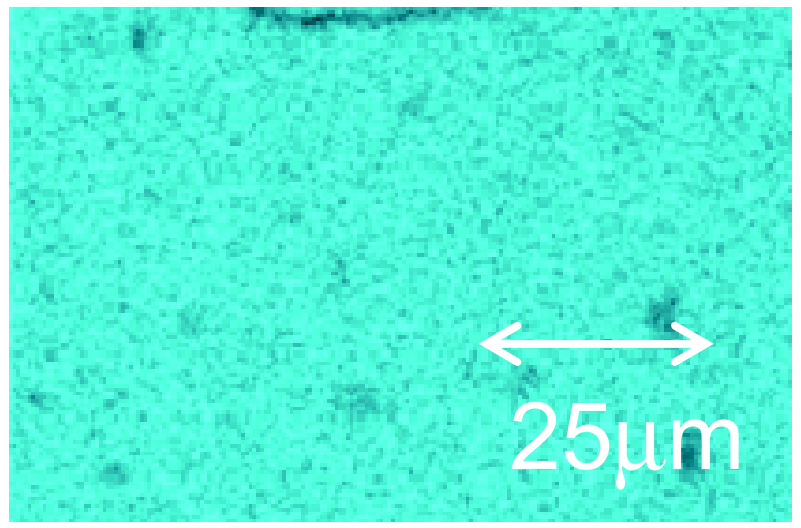

Black: no signal

(c) iron (Fe) Kα emission

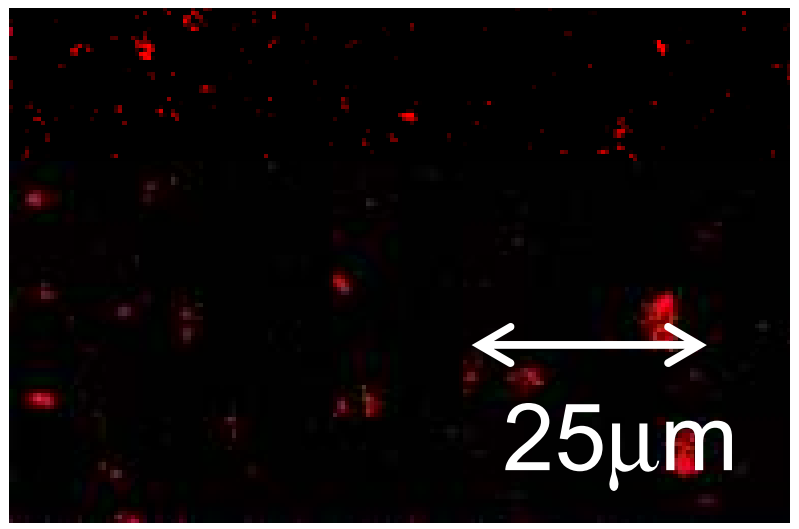

Black: no signal

(d)

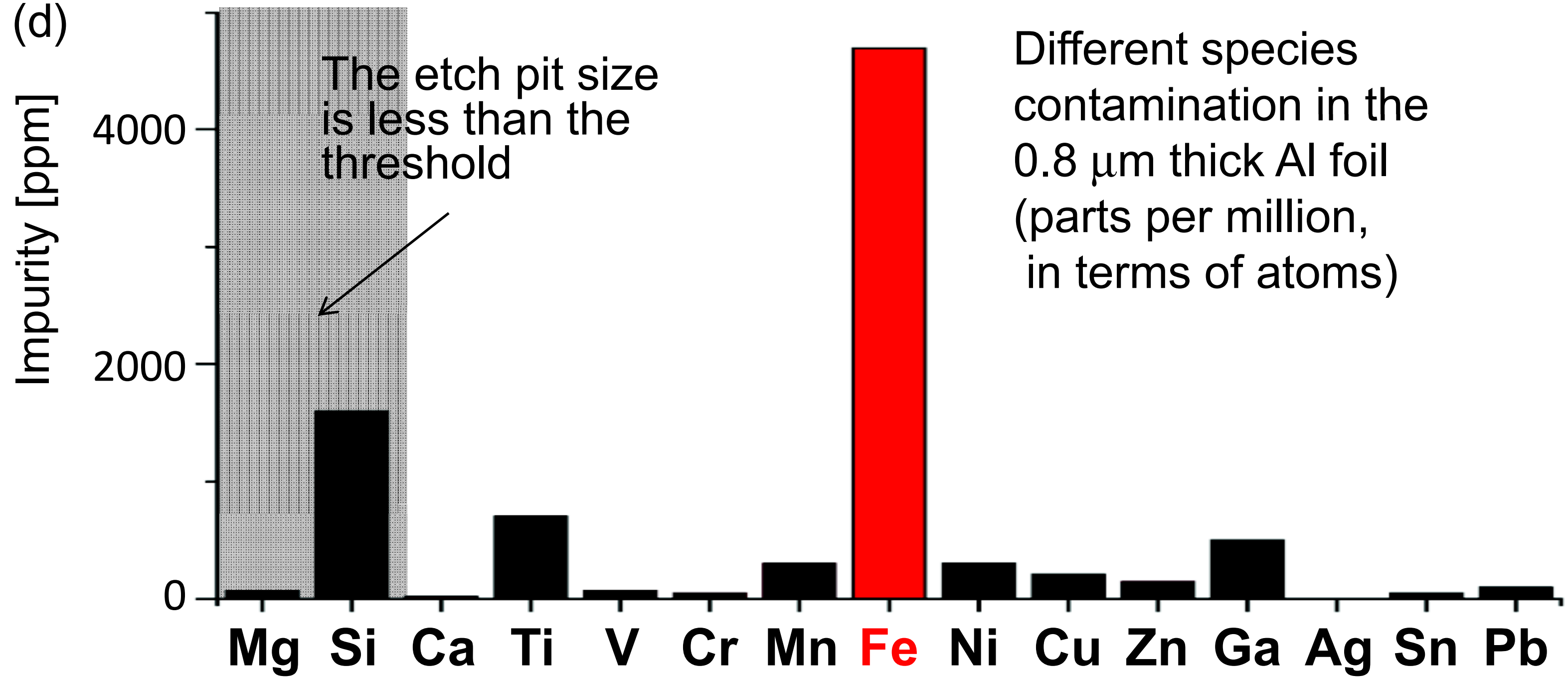

Different species contamination in the 0.8 μm thick Al foil (parts per million, in terms of atoms)

Fig.SI_3

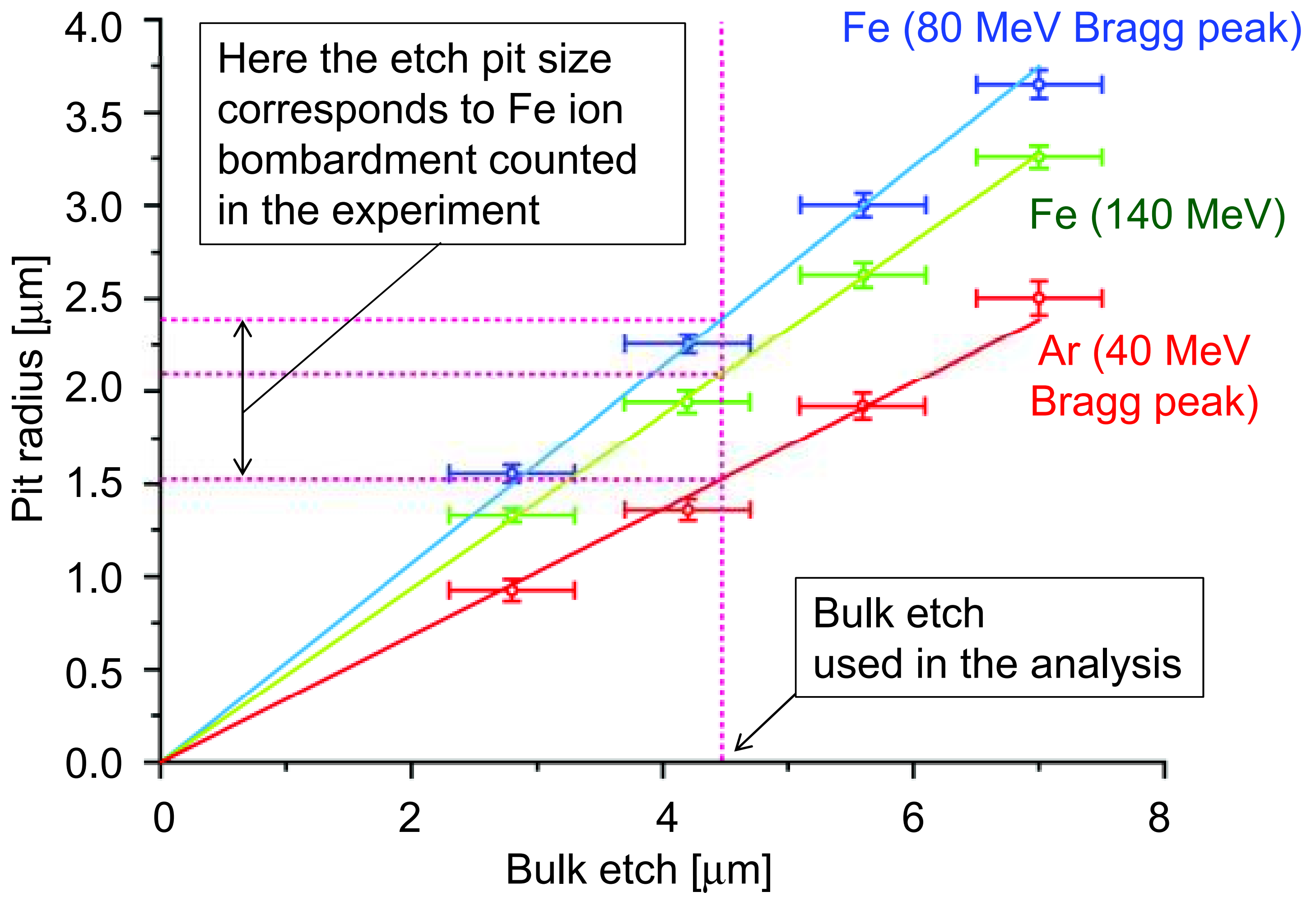

Here the etch pit size corresponds to Fe ion bombardment counted in the experiment
Fe (80 MeV Bragg peak)
Fe (140 MeV)
Ar (40 MeV Bragg peak)
Bulk etch used in the analysis
Pit radius [μm]
4.0
3.5
3.0
2.5
2.0
1.5
1.0
0.5
0.0
0
2
4
6
8
Bulk etch [μm]

Fig.SI_4

Energy of $^{248}$Cm ions in [MeV/u]
as a function of the laser pulse energy, $\mathcal{E}_{las}$ [J],
and target thickness, $\ell$ [μm].

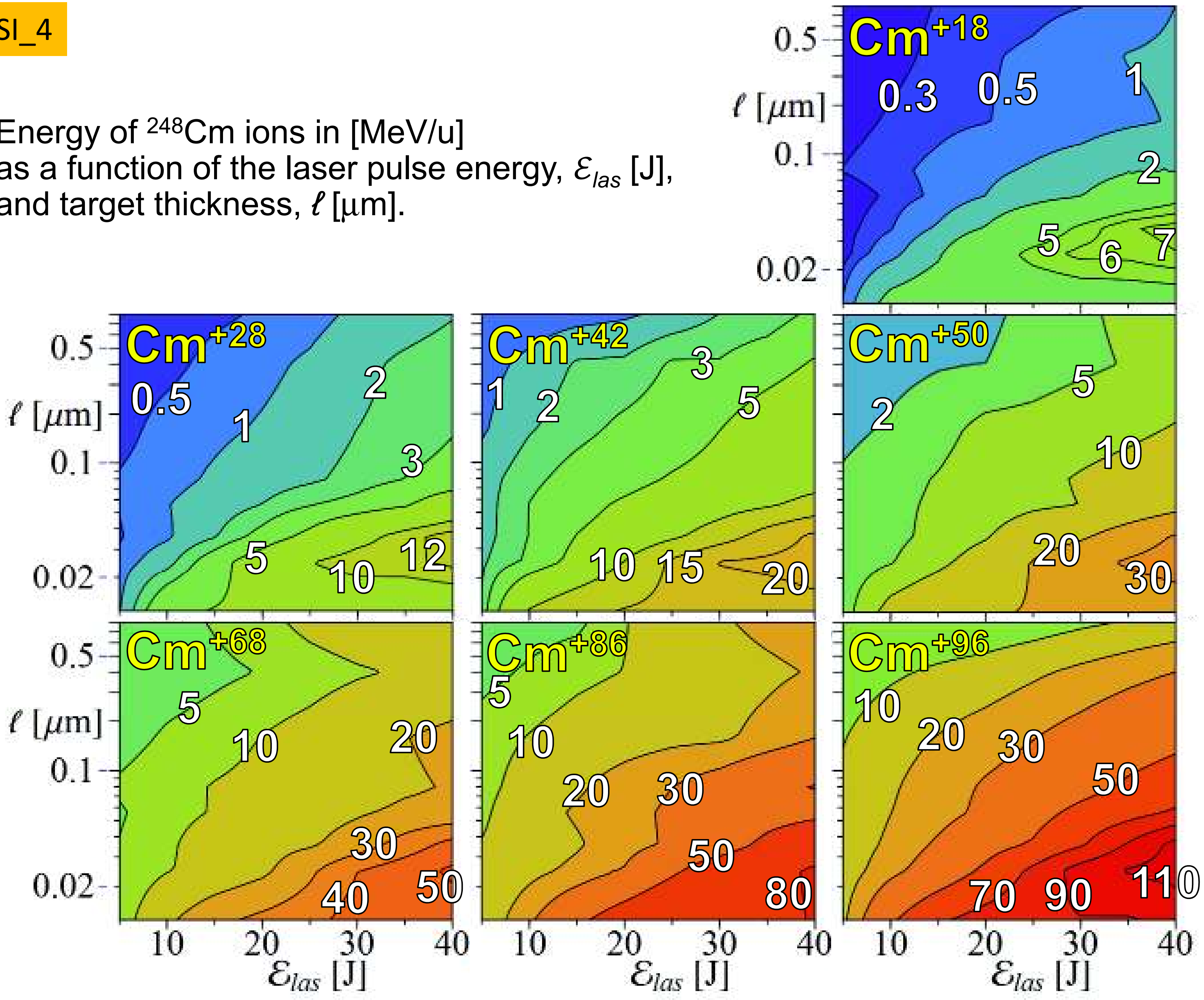

Fig.SI_5

Number of $^{248}Cm$ ions with energy ≥ $\mathcal{E}$ [MeV/u], as a function of the laser pulse energy and target thickness.

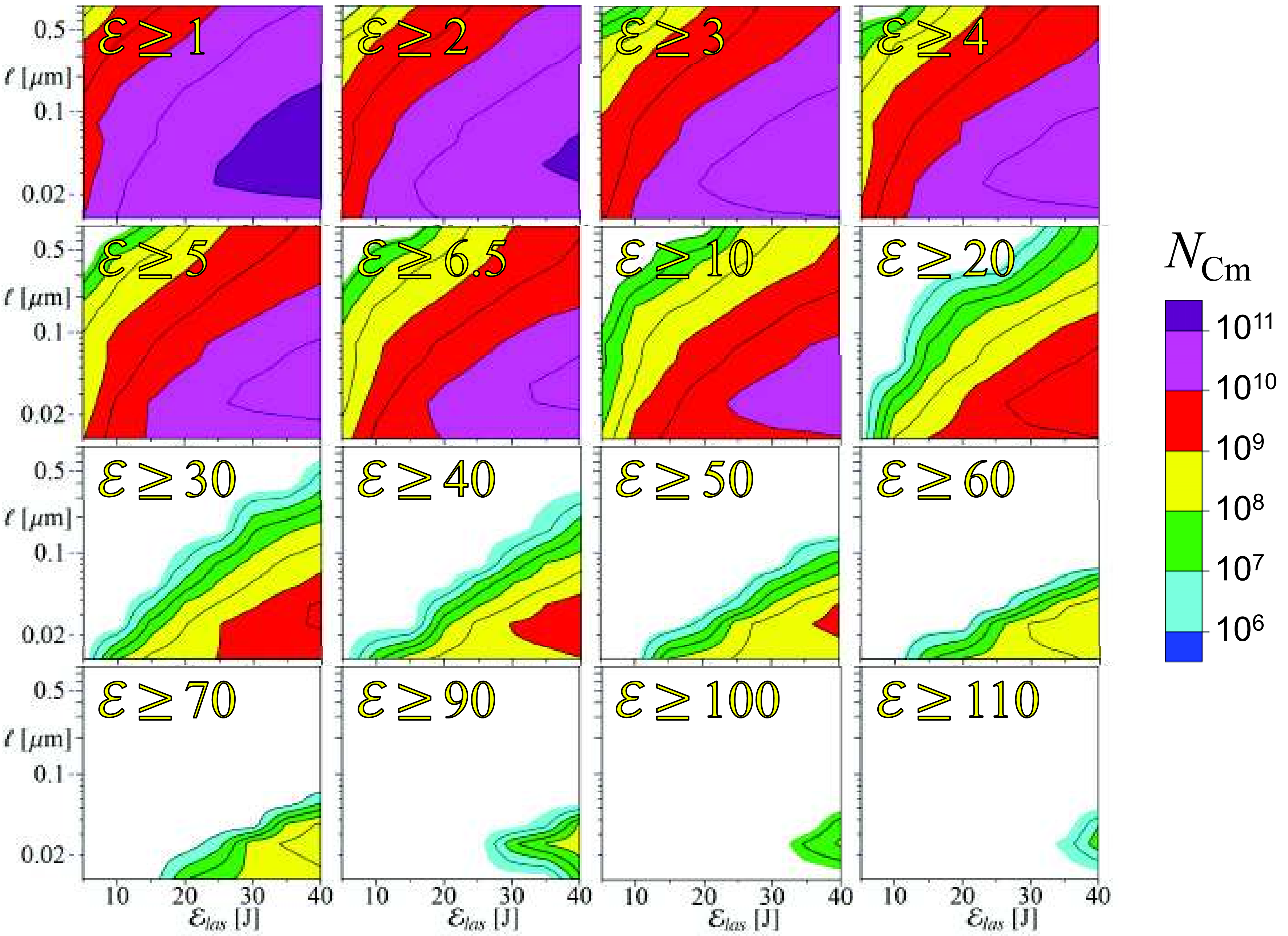

Fig.SI_6

Laser pulse interaction with a curium-248 foil.

Laser pulse energy
$\mathcal{E}_{las} = 35$ J

Foil thickness
$\ell = 0.8$ μm

The resulting maximum ion energy
$\mathcal{E}_{Cm}^{max} = 22.8$ MeV/u

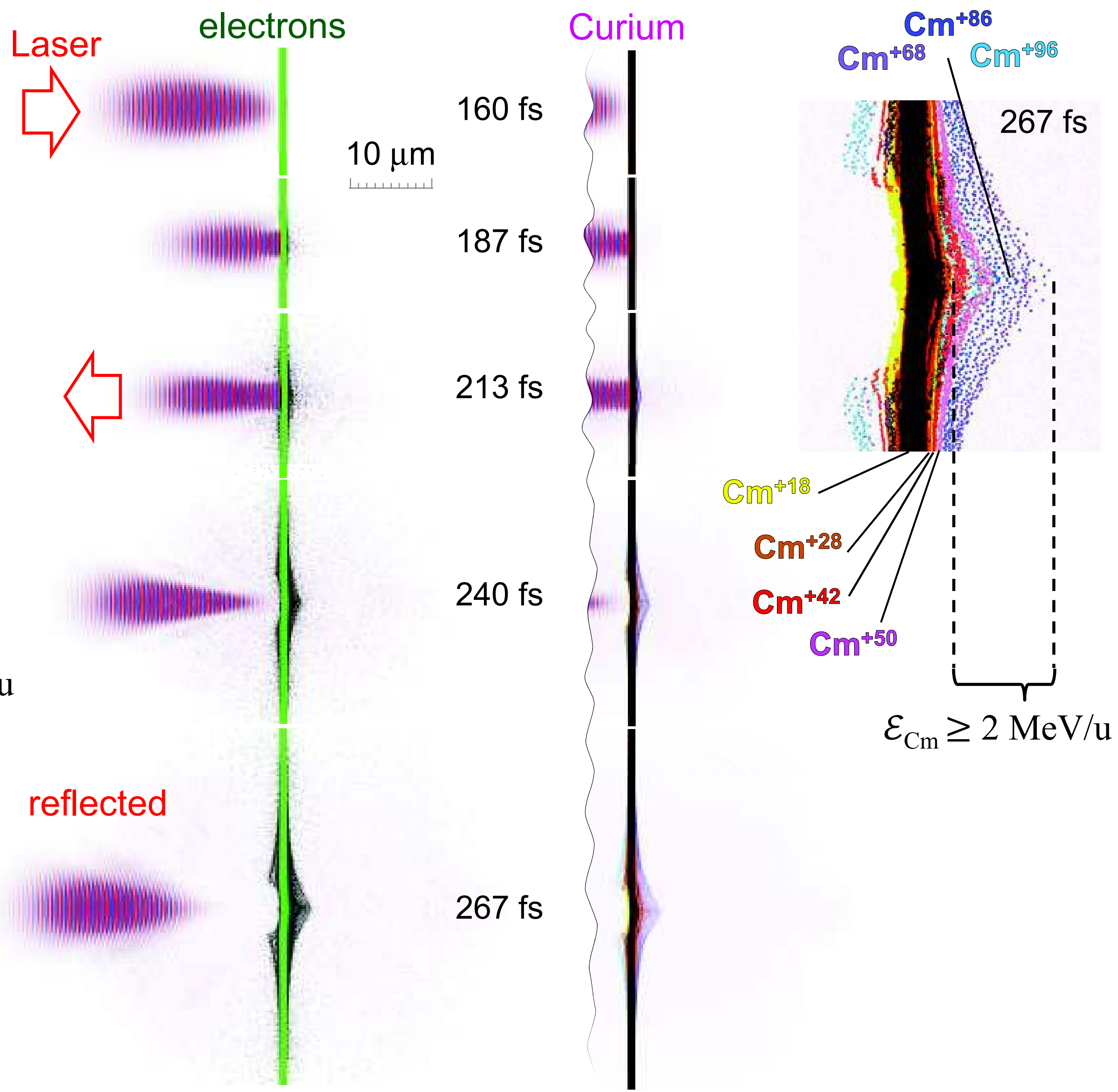

Fig.SI_7

Laser pulse interaction with a curium-248 foil.

Laser pulse energy
$\mathcal{E}_{las} = 40$ J

Foil thickness
$\ell = 25$ nm

The resulting maximum ion energy
$\mathcal{E}_{Cm}^{max} = 117$ MeV/u

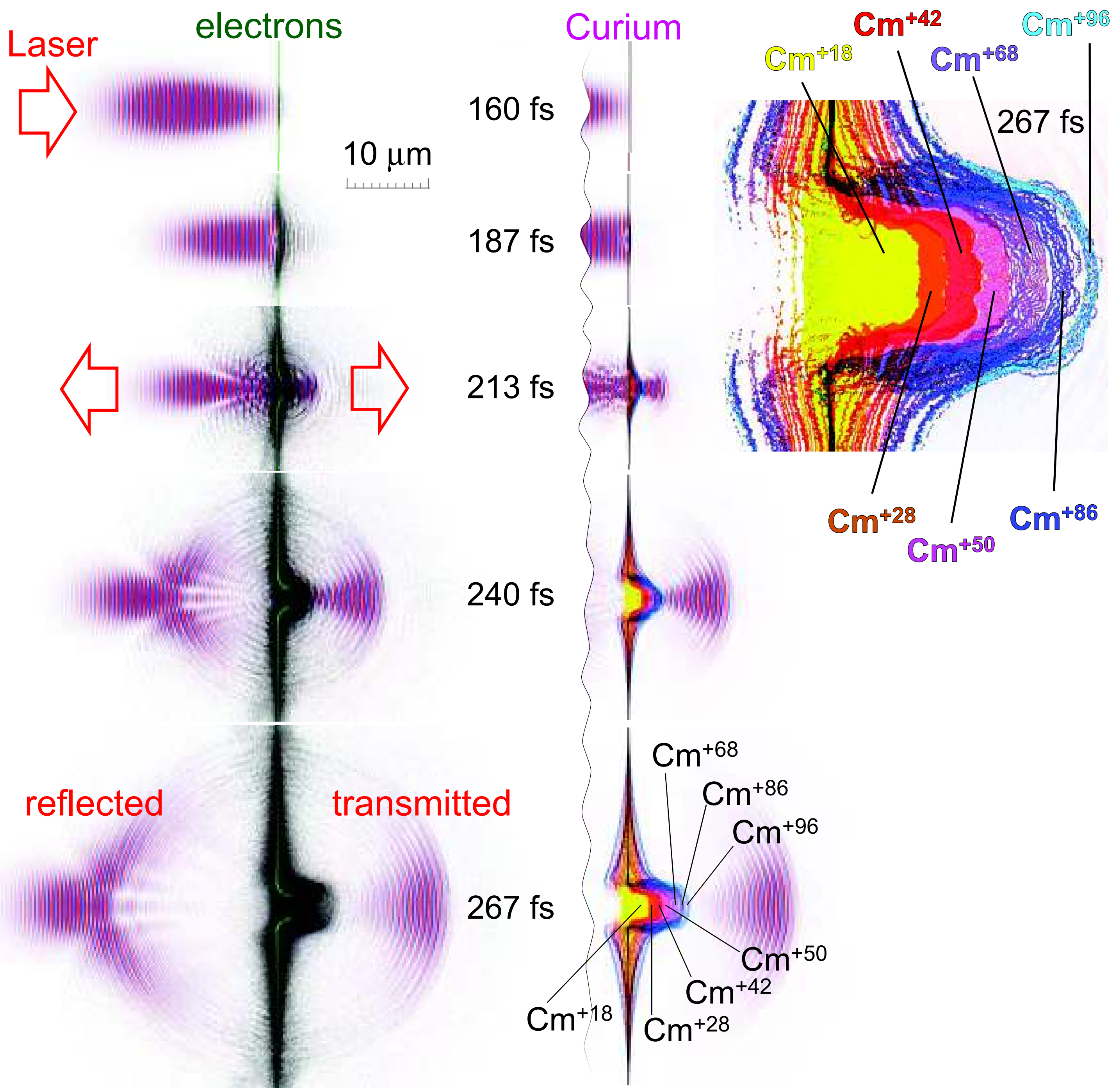

SI_2

Target information.

Images of the 0.8 μm thick Al foil surface produced with scanning electron microscope by

(a) reflected electrons, (b) aluminium (Al) Kα emission, and (c) iron (Fe) Kα emission.

(d) Different species contamination in the 0.8 μm thick Al foil (parts per million, in terms of atoms).

SI_3

The Kapton film response

The Kapton film response is calibrated using a 40 MeV Ar ion beam and 140 MeV and 80 MeV Fe ion beams from the conventional accelerator HIMAC in Chiba, Japan. The growth rate of the pit size in the etching process linearly depends on the bulk etch rate, which is defined as the bulk thickness removed in the etching process.

SI_4

Result of multi-parametric simulations(1)

Maximum ion energies, $\mathcal{E}$Cm, corresponding to different Q/M ratios

Idealized laser pulse and a foil target is used.

SI_5

Result of multi-parametric simulations(2)

The number of ions accelerated to an energy above certain threshold .

Idealized laser pulse and a foil target is used.

SI_6

Result of multi-parametric simulations(3)

Laser pulse interaction with a curium-248 foil in the case of laser pulse energy of 35 J and foil thickness of 0.8μm.

SI_7

Result of multi-parametric simulations(4)

Laser pulse interaction with a curium-248 foil in the case of laser pulse energy of 40 J and foil thickness of 25nm.